\documentclass[twocolumn,showkeys]{revtex4}
\usepackage[utf8]{inputenc}
\usepackage{graphicx}

\begin{document}

\title{Grey solitons in the ultracold fermions at the full spin polarization}

\author{Pavel A. Andreev}
\email{andreevpa@physics.msu.ru}
\affiliation{Faculty of physics, Lomonosov Moscow State University, Moscow, Russian Federation, 119991.}
\affiliation{Peoples Friendship University of Russia (RUDN University), 6 Miklukho-Maklaya Street, Moscow, 117198, Russian Federation}

\date{\today}

\begin{abstract}
A minimal coupling quantum hydrodynamic model of spin-1/2 fermions at the full spin polarization
corresponding to a nonlinear Schrodinger equation is considered.
The nonlinearity is primarily caused by the Fermi pressure.
It provides an effective repulsion between fermions.
However, there is the additional contribution of the short-range interaction appearing in the third order by the interaction radius.
It leads to the modification of the pressure contribution.
Solitons are considered for the infinite medium with no restriction on the amplitude of the wave.
The Fermi pressure leads to the soliton in form of the area of decreased concentration.
However, the center of solution corresponding to the area of minimal concentration has nonzero value of concentration.
Therefore, the grey soliton is found.
Soliton exist if the speed of its propagation is below the Fermi velocity.
\end{abstract}



\maketitle


\section{Introduction}

Solitons are fundamental nonlinear structures existing in various physical systems including the quantum gases
\cite{Alotaibi PRA 17}, \cite{Andreev MPL B 12}, \cite{Wang NJP 14}, \cite{Andreev 2009}, \cite{Katsimiga NJP 17},
\cite{Shamailov PRA 19}, \cite{Syrwid PRA 17}, \cite{Syrwid PRA 18}, \cite{Wang Dai PRA 18}, \cite{Zhao PRA 20}.
Vortexes, shock waves, and skyrmions are among main nonlinear objects considered in atomic quantum gases.
Solitons and other nonlinear phenomena are well studied in the bosonic atoms experimentally and theoretically.
The theoretical approach is mainly based on the mean-field nonlinear Schrodinger equation called the Gross-Pitaevskii equation
which describes the bright and dark solitons in the Bose-Einstein condensates (BECs),
where the nonlinearity is caused by the interparticle interaction \cite{Dalfovo RMP 99}.
The form of solitons is related to the sign of interaction,
so the bright and dark solitons appears in the attractive and repulsive BECs, correspondingly.
For instance, the experimental study of the dark solitons in BECs is presented in Ref. \cite{Becker Nat Ph 08}.
Majority of study of fermionic gases are focused on the superfluid phase or BCS state (Bardeen-Cooper-Schrieffer state),
where pairs of fermionic atoms with opposite spins and momentum form the Cooper pairs and demonstrate the boson-like behavior \cite{Giorgini RMP 08}.
The dark solitons in superfluid Fermi gases are considered in Refs. \cite{Antezza PRA 07}, \cite{Scott PRL 11}, \cite{Liao PRA 11},
particularly, the Bogoliubov-de Gennes equations are used in \cite{Antezza PRA 07}.
A heavy soliton in a fermionic superfluid is experimentally observed in Ref. \cite{Yefsah Nat 13}.
Solitons in the superfluid Fermi gases are considered in terms of the nonlocal generalization of the Ginzburg-Landau model \cite{Barkman PRR 20},
following Ref. \cite{Buzdin PLA 97}.
Here, the basic and fundamental solitons are considered in degenerate fermionic atoms with the full spin polarization.
The spin-1/2 atoms are chosen,
but the same analysis is correct for the fermions with higher spins.
The presented theoretical work is based on the quantum hydrodynamics
which is straightforwardly
derived from the microscopic many-particle Schrodinger equation (from the full quantum theory).
The minimal coupling model of fermions is composed of two hydrodynamic equations:
the continuity and Euler equation.
These hydrodynamic equations allow to obtain the corresponding nonlinear Schrodinger equation,
where the nonlinearity is mainly caused by the Fermi pressure.
It is nonlinearity of fractional degree $7/3$.
However, the interaction between fermions gives the additional nonlinearity \cite{Andreev 2001}.

Unpolarized fermions  are mostly discussed in literature regime for fermions \cite{Kulkarni PRA 12}, \cite{Babadi PRA 12}, \cite{Andreev 1912}.
If we have system of spin-1/2 Fermi atoms with equal population of the spin-up state and the spin-down state
we can observe interesting phases of matter.
They are the Bose-Einstein condensate of molecules, crossover superfluid, and the BCS state.
The spin orbit coupling can also be engineered in the unpolarized fermions.

If we consider bosons being in the Bose-Einstein condensate state
it can be described by the Gross-Pitaevskii equation.
The interaction appears in the first order by the interaction radius in term of the hydrodynamic derivation of the Gross-Pitaevskii equation.
Or it can be interpreted as the s-wave scattering in terms of the scattering theory.
For the polarized fermions, there is no contribution of the interaction in the first order by the interaction radius (FOIR),
due to the antisymmetry of the wave function.
Hence, main selfaction of the fermion fluid comes from the Fermi pressure.
However, the additional contribution of the interaction appearing in the third order by the interaction radius (TOIR) can be derived \cite{Andreev 2001}.
This contribution can be interpreted via the p-wave scattering in terms of the scattering theory
\cite{Parker PRA 12}, \cite{Roth PRA 01}, \cite{Roth PRA 02}.

The spin polarized fermions is the system where all fermions occupy the single spin state.
This systems shows rather avaricious phase.
Nevertheless, it also demonstrates some interesting fundamental nonlinear phenomena.
To some extend the system of  polarized degenerate fermions can be describe by the effective macroscopic single particle wave function $\Phi(\textbf{r},t)$.
This possibility follows from the quantum hydrodynamic equations restricted by the particle density and the momentum density evolution.
However, complete description of polarized fermions requires the momentum current evolution,
which is the kinetic pressure of polarized fermions.
The pressure evolution equation gives more accurate value of the speed of sound \cite{Andreev 2001}, \cite{Andreev 1912}, \cite{Tokatly PRB 99}.
Nevertheless, the minimal coupling model based on the particle density and the momentum density shows good qualitative description of fermions.

This paper is organized as follows.
In Sec. II the quantum hydrodynamics is presented in two regimes: the mean-field approximation and up to the TOIR approximation.
In Sec. III solution of the hydrodynamic equations in the one-dimensional regime in the form of the grey soliton
is obtained by the Sagdeev potential method.
In Sec. IV a brief summary of obtained results is presented.

\section{Quantum hydrodynamic equations}

Here, we present two quantum hydrodynamic models for the degenerate fermions being in the same spin state (the regime of the full spin polarization).
The first model is obtained in the FOIR approximation,
where the interaction gives the zero contribution.
The second model contains the contribution of the interaction in the TOIR approximation.

\subsection{First order by the interaction radius: A minimal coupling hydrodynamic model for the full spin polarization}

Nonzero contribution of the interaction in the first order by the interaction radius exists for nonpolarized or the partially polarized systems of fermions.
However, the fully polarized fermions have zero contribution of interaction in this case.

In all regimes we have same form of the continuity equation:
\begin{equation}\label{DsolF cont eq via vel spin up} \partial_{t}n+\nabla\cdot (n\textbf{v})=0, \end{equation}

In the FOIR approximation we also have the Euler (momentum balance) equation
$$ mn(\partial_{t} +\textbf{v}\cdot\nabla)\textbf{v}
-\frac{\hbar^{2}}{2m}n\nabla\frac{\triangle\sqrt{n}}{\sqrt{n}}$$
\begin{equation}\label{DsolF Euler FOIR}
+\nabla p=-n\nabla V_{ext},\end{equation}
where $p$ is the Fermi pressure
\begin{equation}\label{DsolF Fermi pressure} p= \frac{(6\pi^{2})^{\frac{2}{3}}\hbar^{2}n^{\frac{5}{3}}}{5m^{2}}.
\end{equation}

Minimal coupling assumes the application of the continuity and Euler equation with no account of the pressure evolution,
but application of the equation of state for the reduction of the pressure evolution to the concentration evolution.

Equations (\ref{DsolF cont eq via vel spin up})-(\ref{DsolF Fermi pressure})
correspond to the nonlinear Schrodinger equation for fermions at the potential velocity field \cite{Andreev 2001}:
\begin{equation}\label{DsolF NLSE first appearence} \imath\hbar\partial_{t}\Phi
=\biggl(-\frac{\hbar^{2}\nabla^{2}}{2m}+\frac{(6\pi^{2}n)^{\frac{2}{3}}\hbar^{2}}{2m}+V_{ext}\biggr)\Phi,\end{equation}
where $n=\mid \Phi\mid^{2}$.
The effective macroscopic wave function $\Phi$ is defined
via the hydrodynamic wave functions $n(\textbf{r},t)$ and $\textbf{v}(\textbf{r},t)$:
\begin{equation}\label{DsolF def Phi} \Phi(\textbf{r},t)=\sqrt{n} e^{\imath m\phi/\hbar},\end{equation}
where $\textbf{v}=\nabla\phi$.
The contribution of the Fermi pressure (\ref{DsolF Fermi pressure}) is presented by the second term
on the right-hand side of equation (\ref{DsolF NLSE first appearence}).

Equation similar to NLSE (\ref{DsolF NLSE first appearence}) are used in literature
\cite{Adhikari PRA 04},
\cite{Adhikari PRA05}, \cite{Adhikari JPB05}, \cite{Adhikari NJP06}, \cite{Bludov PRA06}, 
\cite{Rizzi PRA08}, \cite{Maruyama PRA08}, \cite{Karpiuk PRA06}.
Different forms have different justifications.
Equation (\ref{DsolF NLSE first appearence}) is justified via the quantum hydrodynamics.
Moreover, the partial or full spin polarization is not included there.

Equations (\ref{DsolF cont eq via vel spin up}) and (\ref{DsolF Euler FOIR}) are applied below to consider the possibility of solitons in the systems of neutral atomic degenerate fermions.
To complete the description of model
we present the hydrodynamics containing the contribution of the interaction between fermions with the same spin polarization.
Absence of the interaction in equations (\ref{DsolF Euler FOIR}) and (\ref{DsolF NLSE first appearence}) shows
that the equilibrium condition cannot be reached in such systems.
However, we have interaction between fermions
which is presented below.
It provides the additional transfer of the momentum and a mechanism of reaching of the equilibrium state.

\subsection{Hydrodynamic equations and nonlinear Schrodinger equation for fermions with the interaction included up to the TOIR approximation}

Derivation of the macroscopic equations by the many-particle quantum hydrodynamics method
\cite{Maksimov QHM 99}, \cite{Andreev 2005}, \cite{MaksimovTMP 2001}, \cite{Andreev 2007}, \cite{Andreev LP 19} shows that
the hydrodynamic equations appear first.
Next, in some simplified regimes the nonlinear Schrodinger equation can be found \cite{Andreev 2001}, \cite{Maksimov QHM 99}.

The nonlinear Schrodinger equation can be derived in the chosen approximation \cite{Andreev 2001}
\begin{equation}\label{DsolF NLSE TOIR} \imath\hbar\partial_{t}\Phi
=\biggl(-\frac{\hbar^{2}\nabla^{2}}{2m}+\frac{(6\pi^{2}n)^{\frac{2}{3}}\hbar^{2}}{2m}+V_{ext} -\frac{4}{5}g_{2}(6\pi^{2})^{\frac{2}{3}}n^{\frac{5}{3}}\biggr)\Phi,\end{equation}
the additional term caused by the interaction is the last term in equation (\ref{DsolF NLSE TOIR}).
The additional term is obtained in the TOIR approximation.
It contains the interaction constant $g_{2}$ which is defined via the potential of interatomic interaction $U$:
\begin{equation}\label{DsolF g 2 def} g_{2}=\int r^{2}U(r)d\textbf{r}. \end{equation}
The positive interaction constant decreases the pressure.
However, the model is obtained in the weak interaction limit.
So, the contribution of interaction should be small in compare with the Fermi pressure.

Let us present the corresponding hydrodynamic equations.
The continuity equation has same form (\ref{DsolF cont eq via vel spin up}).
The Euler equation contains the additional term
$$ mn(\partial_{t} +\textbf{v}\cdot\nabla)\textbf{v}
-\frac{\hbar^{2}}{2m}n\nabla\frac{\triangle\sqrt{n}}{\sqrt{n}}$$
\begin{equation}\label{DsolF Euler TOIR}
+\nabla p=-n\nabla V_{ext}+\frac{5m^{2}}{2\hbar^{2}}g_{2}\nabla (n p),\end{equation}
which is the last term in equation (\ref{DsolF Euler TOIR}).
The interaction term appear via the kinetic pressure $p$ \cite{Andreev 2001}.
The gradient of the kinetic pressure itself also presented by the last term on the left-hand side of equation (\ref{DsolF Euler TOIR}).
In this paper we use the equation of state in form of the Fermi pressure (\ref{DsolF Fermi pressure}).
Hence, the Euler equation (\ref{DsolF Euler TOIR}) is truncated
and its final form corresponds to the nonlinear Schrodinger equation (\ref{DsolF NLSE TOIR}).

The model presented above is obtained for the fermions with the full spin polarization.
Hydrodynamic model of degenerate spin-1/2 fermions with the partial spin polarization
in the mean-field approximation for the interaction between fermions with different spin projections \cite{Andreev LPL 18}.

\section{Large amplitude grey solitons}

Let us present the analysis of equations for the solitons obtained up to the TOIR approximation.
Let us consider  solitons in uniform infinite medium.
Hence, the symmetry of the system allows to consider the nonlinear waves with the plane wave front.
The wave appears as the one dimensional solution.
We consider the wave propagation in the arbitrary direction and choose the cartesian coordinates with axis $Ox$ in the direction of the wave propagation.
We seek the stationary solutions of the nonlinear equations.
We consider the steady state in the comoving frame.
Hence, all hydrodynamic functions depend on $\eta=x-u t$ and $u$,
where the parameter $u$ is the constant velocity of the nonlinear solution.
The perturbations vanish at $\eta\rightarrow\pm\infty$.

\subsection{One dimensional limit of hydrodynamic equations}

For the nonlinear plane waves we have the following simplified continuity equation
\begin{equation}\label{DsolF cont simple 1D} -u\partial_{\eta}n+\partial_{\eta}(nv^{x})=0, \end{equation}
where
the time derivative $\partial_{t}$ is replaced by $-u\partial_{\eta}$ in accordance with the variable $\eta$ introduced for the stationary solution.
Similar simplification is made for the Euler equation
$$-umn\partial_{\eta}v^{x} +mnv^{x}\partial_{\eta}v^{x}
-\frac{\hbar^{2}}{2m}n\partial_{\eta}\frac{\partial_{\eta}^{2}\sqrt{n}}{\sqrt{n}}$$
\begin{equation}\label{DsolF Euler simple 1D}
=-\frac{(6\pi^{2})^{2/3}\hbar^{2}}{2m^{2}}n\partial_{\eta}n^{2/3}
+g_{2}\frac{4(6\pi^{2})^{2/3}}{5}n\partial_{\eta}n^{5/3},
\end{equation}
where the terms placed on the right-hand side are represented via construction $n\partial_{\eta}n^{a}$
useful for the further transformations, with $a$ is the arbitrary degree.

The one dimensional continuity equation (\ref{DsolF cont simple 1D}) can be integrated
\begin{equation}\label{DsolF cont simple 1D integrated} n(v^{x}-u)=-un_{0}, \end{equation}
where the boundary conditions $n(\eta\rightarrow\pm\infty)=n_{0}$,
and $v^{x}(\eta\rightarrow\pm\infty)=0$ are used.
Equation (\ref{DsolF cont simple 1D integrated}) allows to express the velocity field via the concentration
\begin{equation}\label{DsolF velocity via n} v^{x}=\frac{u(n-n_{0})}{n}. \end{equation}
All terms in the Euler equation (\ref{DsolF Euler simple 1D}) are proportional to the concentration,
so we can drop it.
Next, the Euler equation can be integrated.
As the result we find
$$m\biggl(\frac{1}{2}v_{x}^{2} -uv^{x}\biggr)
-\frac{\hbar^{2}}{2m}\frac{\partial_{\eta}^{2}\sqrt{n}}{\sqrt{n}}$$
$$+\frac{(6\pi^{2})^{2/3}\hbar^{2}}{2m^{2}}n^{2/3}
-g_{2}\frac{4(6\pi^{2})^{2/3}}{5}n^{5/3}$$
\begin{equation}\label{DsolF Euler simple 1D integrated}
=\frac{(6\pi^{2})^{2/3}\hbar^{2}}{2m^{2}}n_{0}^{2/3}
-g_{2}\frac{4(6\pi^{2})^{2/3}}{5}n_{0}^{5/3},
\end{equation}
where all terms existing in equation (\ref{DsolF Euler simple 1D}) are placed on the left-hand side
while the right-hand side contains the result of application of the boundary conditions.

\begin{figure}
\includegraphics[width=8cm,angle=0]{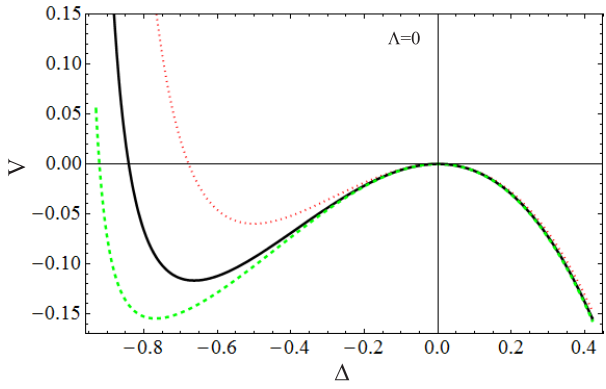}
\caption{\label{DsolF Fig 01} The Sagdeev potential $V=V(\Delta)$
(\ref{DsolF Sagdeev potential FOIR}) is demonstrated for the noninteracting limit
for three different values of the dimensionless speed of the soliton
$\alpha=0.2$ (the upper red dotted line),
$\alpha=0.1$ (the middle black continuous line),
$\alpha=0.05$ (the lower green dashed line).
}
\end{figure}

We substitute the velocity field (\ref{DsolF velocity via n}) in the integrated Euler equation (\ref{DsolF Euler simple 1D integrated}).
Moreover, we see that equation (\ref{DsolF Euler simple 1D integrated}) contains the second derivative on $\sqrt{n}$.
It shows that we should find solution relatively $\sqrt{n}$.
Equation (\ref{DsolF Euler simple 1D integrated}) can be integrated to obtain the "energy integral" in the following manner
\begin{equation}\label{DsolF final eq for n or phi OT}
\frac{1}{2}(\partial_{\eta}\sqrt{n})^{2} +\tilde{V}_{eff}(\sqrt{n})=0,\end{equation}
where the first term can be considered as the effective kinetic energy of soliton,
while $\tilde{V}_{eff}(\sqrt{n})$ is the effective potential energy called the Sagdeev potential
\cite{Schamel PF 77}, \cite{Witt PF 83}, \cite{Mamun PRE 97}, \cite{Shah PoP 10},
\cite{Marklund PRE 07}, \cite{Akbari-Moghanjoughi PP 17}.
The Sagdeev potential $\tilde{V}_{eff}(\sqrt{n})$ appears in the following form
$$\tilde{V}_{eff}(\sqrt{n})=
\frac{1}{2}(6\pi^{2})^{2/3}\biggl(1+\frac{8mg_{2}}{5\hbar^{2}}n_{0}\biggr)(n-n_{0})
$$
$$+\frac{m^{2}u^{2}}{2\hbar^{2}}\biggl(n+\frac{n_{0}^{2}}{n}-2n_{0}\biggr)
-\frac{3}{10}(6\pi^{2})^{\frac{2}{3}}\biggl(n^{\frac{5}{3}}+\frac{mg_{2}}{\hbar^{2}}n^{\frac{8}{3}}\biggr)$$
\begin{equation}\label{DsolF Sagdeev potential}
-\frac{3}{10}(6\pi^{2})^{2/3}\biggl(n_{0}^{5/3}+\frac{mg_{2}}{\hbar^{2}}n_{0}^{8/3}\biggr).\end{equation}

Equations can be solved for parameter $\sqrt{n}$, but the traditional form of the presentation of the results including
the zero value of the effective potential and its first derivative on parameter $\sqrt{n}$
requires to consider dependence on $\sqrt{n}-\sqrt{n}_{0}$.
Let us to choose the dimensionless form of the chosen parameter $\Delta\equiv(\sqrt{n}-\sqrt{n}_{0})/\sqrt{n}_{0}$ for the further analysis.
Moreover, the coordinate in the comoving frame $\eta$ can be presented in the dimensionless form as well $\xi=\eta n_{0}^{1/3}$.

\begin{figure}
\includegraphics[width=8cm,angle=0]{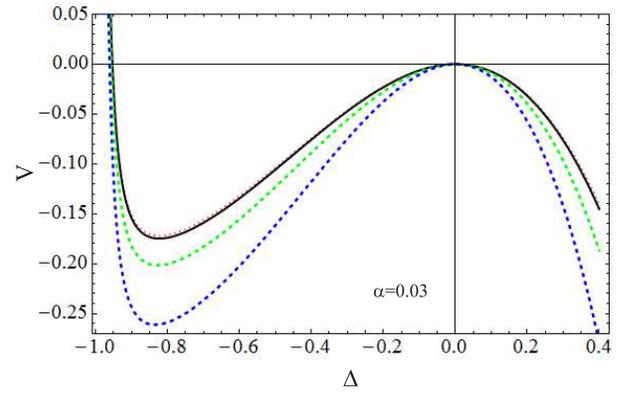}
\caption{\label{DsolF Fig 02}The Sagdeev potential $V=V(\Delta)$ (\ref{DsolF Sagdeev potential TOIR}) is demonstrated
at the account of the interaction up to TOIR approximation.
The Sagdeev potential $V=V(\Delta)$ is demonstrated for the fixed speed of perturbation $\alpha=0.03$
for different values of the dimensionless interaction constant
$\Lambda=0$ (the upper red dotted line),
$\Lambda=0.01$ (the second from above black continuous line),
$\Lambda=0.1$ (the third from above green dashed line),
$\Lambda=0.3$ (the lower blue dashed line).
}
\end{figure}

\subsection{Grey soliton in the mean-field approximation}

The mean-field approximation corresponds to the FOIR limit of the hydrodynamic equations.
Formally, it can be obtained from equations (\ref{DsolF final eq for n or phi OT}) and (\ref{DsolF Sagdeev potential}) at $g_{2}=0$.
we discuss the Sagdeev potential in the dimensionless form.

The dimensionless form of the Sagdeev potential
$V(\Delta)\equiv\tilde{V}_{eff}(\sqrt{n})n_{0}^{-5/3}$
presented in the mean-field approximation has the following form
$$V(\Delta)=\frac{1}{2}\alpha^{2}\biggl[ (\Delta+1)^{2}+\frac{1}{(\Delta+1)^{2}}\biggr]$$
\begin{equation}\label{DsolF Sagdeev potential FOIR}
+\frac{1}{2}(\Delta+1)^{2}
-\frac{3}{10}(\Delta+1)^{10/3}
-\biggl[\alpha^{2}+\frac{1}{5}\biggr]
,\end{equation}
where $\alpha\equiv u/v_{Fe}$.
The dimensionless Sagdeev potential (\ref{DsolF Sagdeev potential FOIR}) is a part of the following dimensionless equation
$(1/2)(\partial_{\xi}\Delta)^{2} +V(\Delta)=0$.

\begin{figure}
\includegraphics[width=8cm,angle=0]{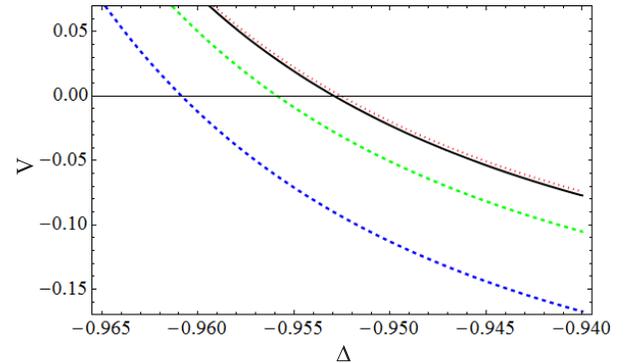}
\caption{\label{DsolF Fig 03}
A part of the Sagdeev potential $V=V(\Delta)$ (\ref{DsolF Sagdeev potential TOIR}) is demonstrated.
It is a part of Fig. (\ref{DsolF Fig 02})
which corresponds to the point of crossing of the Sagdeev potential of the line of zero potential.
}
\end{figure}

\begin{figure}
\includegraphics[width=8cm,angle=0]{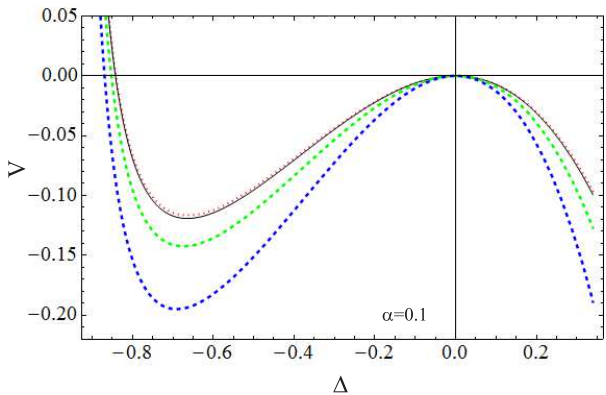}
\caption{\label{DsolF Fig 04} This figure is similar to Fig. (\ref{DsolF Fig 02}),
but this figure is obtained for the larger speed of the soliton.
The Sagdeev potential $V=V(\Delta)$ (\ref{DsolF Sagdeev potential TOIR}) is demonstrated
at the account of the interaction up to TOIR approximation.
The Sagdeev potential $V=V(\Delta)$ is demonstrated for the fixed speed of perturbation $\alpha=0.1$
for different values of the dimensionless interaction constant
$\Lambda=0$ (the upper red dotted line),
$\Lambda=0.01$ (the second from above black continuous line),
$\Lambda=0.1$ (the third from above green dashed line),
$\Lambda=0.3$ (the lower blue dashed line).
}
\end{figure}

Fig. (\ref{DsolF Fig 01}) shows the single illustration of the Sagdeev potential in the mean-field regime (\ref{DsolF Sagdeev potential FOIR}).
Value $\Delta_{0}\neq0$ corresponding to $V(\Delta_{0})=0$ shows the amplitude of the soliton .
First, we see that
$\Delta_{0}$ is negative.
Hence, there is the decrease of concentration in the soliton $n<n_{0}$.
However, $\Delta_{0}$ does not reach value $-1$.
Consequently,  the concentration of particles in the soliton is always nonzero $n>0$.
The soliton appears as the area of decreased concentration,
which is above the zero value at the center of soliton.
Such solitons are called the gray soliton.
While the dark soliton is the limiting case of the grey soliton with the zero concentration in its center.

Fig. (\ref{DsolF Fig 01}) shows that
the increase of the speed of the soliton propagation up to the Fermi velocity decreases the amplitude of the soliton
down to the zero value at $\alpha\approx0.6$.
Moreover, no solution exists at $\alpha>0.6$.
Obtained behavior shows similarity to the dark soliuton in the BECs,
where the speed of soliton propagation is limited by the Landau critical velocity \cite{Busch PRL 00}.

Dimensionless velocity $\alpha$ is the single parameter in the mean-field approximation.
This dependence is discussed.
Further analysis of the properties of soliton can be made in the TOIR approximation.


Presented here soliton solution for the spin polarized fermions.
The spin-0 BECs demonstrate two fundamental solitons in the mean-field regime.
Moreover, the spin-0 BEC show the beyond mean-field bright soliton in the repulsive BEC regime.
The boson-boson and boson-fermion mixtures show some additional soliton related effects.
Particularly, the boson-fermion mixture of the spin-0 BEC and spin-polarized spin-1/2 fermions is considered in Ref. \cite{Andreev LP 21},
where focus is made on the modification of properties
of the beyond mean-field bright soliton existing in the repulsive BECs under influence of the fermions.
Hence, there is no direct relation between the grey soliton given here and the fermion part of the relation demonstrated in Ref. \cite{Andreev LP 21}.

\subsection{Generalization of the grey soliton solution up to the TOIR}

Complete expression of the dimensionless form of the Sagdeev potential
$V(\Delta)\equiv\tilde{V}_{eff}(\sqrt{n})n_{0}^{-5/3}$
obtained from the expression (\ref{DsolF Sagdeev potential}) derived up to the TOIR approximation can be written in the following form
$$V(\Delta)=\frac{1}{2}\alpha^{2}\biggl[ (\Delta+1)^{2}+\frac{1}{(\Delta+1)^{2}}\biggr]$$
$$+\frac{1}{2}(\Delta+1)^{2}(1+\Lambda)
-\frac{3}{10}(\Delta+1)^{10/3}$$
\begin{equation}\label{DsolF Sagdeev potential TOIR}
-\frac{3}{16}\Lambda(\Delta+1)^{16/3}
-\biggl[\alpha^{2}+\frac{1}{2}(1+\Lambda)-\frac{3}{10}-\frac{3}{16}\Lambda\biggr]
.\end{equation}
Additional terms in compare with the FOIR approximation (\ref{DsolF Sagdeev potential FOIR}) are proportional to $\Lambda$.

Contribution of the short-range interaction obtained in the TOIR approximation in the Sagdeev potential is demonstrated
in Figs. (\ref{DsolF Fig 02}), (\ref{DsolF Fig 03}), and (\ref{DsolF Fig 04}).
Figs. (\ref{DsolF Fig 02}) and (\ref{DsolF Fig 03}) are obtained for the relatively small speed of soliton $\alpha=0.03$.
In this case, there is small modification of the amplitude of soliton under the change of the interaction constant.
So, this modification is demonstrated in Fig. (\ref{DsolF Fig 03}).
Fig. (\ref{DsolF Fig 04}) presents the Sagdeev potential for the same values of the interaction constant,
but it is obtained for the larger velocity $\alpha=0.1$.
The contribution of interaction is larger in this velocity regime.
Further increase of the velocity $\alpha\rightarrow 1$ gives large modification of the amplitude under influence of the interaction.
This limit is not presented in figures
since it is beyond the area of applicability of the model,
which corresponds to the weak interaction regime.

\section{Conclusion}

Grey soliton has been found in the system of weakly interacting fermions being in quantum states with the same spin projection.
It has been found from the quantum hydrodynamic equations corresponding to the nonlinear Schrodinger equation.
The soliton has been found and studied within the Sagdeev potential method.
The form of the Sagdeev potential allows to find the amplitude and width of the soliton.
Particularly, it has been found that the concentration is decreased,
but it does not reach the zero value.
Thus, the soliton is classified as the grey soliton.
The change of the grey soliton parameters at the modification of the speed of the soliton and the interaction constant has been analyzed.
We have concluded that the polarized fermions demonstrate the existence of one kind of soliton,
i.e. the soliton of the partial rarification called the grey soliton.
It is in contrast with the Bose-Einstein condensate,
where two kinds of solitons are possible for the uniform medium.
The forms of solitons for the bosons depend on the sign of the interaction between the Bose atoms.
The dark (bright) soliton corresponds to the repulsive (attractive) interaction.
The degenerate fermions are mainly affected by the Fermi pressure
which provides the effective repulsion.
So the dark/grey solitons is the possible structure.

\section{Acknowledgements}
Work is supported by the Russian Foundation for Basic Research (grant no. 20-02-00476).
This paper has been supported by the RUDN University Strategic Academic Leadership Program.

\section{DATA AVAILABILITY}
Data sharing is not applicable to this article as no new data were
created or analyzed in this study, which is a purely theoretical one.

\end{document}